\documentclass{article}
\usepackage{spconf,amsmath,graphicx, amssymb, mathtools, mdframed, xcolor, tikz, hyperref}
\usepackage{siunitx}
\usetikzlibrary{spy, shapes.geometric, positioning}
\graphicspath{
    {figures/}
}

\DeclareMathOperator{\pois}{\mathcal{P}}
\newcommand{\ytrue}{\mathcal{Y}_{\mathrm{obs}}}
\newcommand{\ysim}{\mathcal{Y}_{\mathrm{sim}}}
\newcommand{\ptrue}{p_{\bar{Y}}}
\newcommand{\psim}{p_Y}
\newcommand{\xtrue}{\bar{x}}
\newcommand{\btrue}{\bar{b}}
\newcommand{\xsim}{x}
\newcommand{\bsim}{b}
\newcommand{\Rn}{\mathbb{R}^n}
\newcommand{\Rm}{\mathbb{R}^m}
\newcommand{\Rnn}{\mathbb{R}^{n\times n}}
\newcommand{\Rmn}{\mathbb{R}^{m\times n}}

\makeatletter
\def\bstctlcite{\@ifnextchar[{\@bstctlcite}{\@bstctlcite[@auxout]}}
\def\@bstctlcite[#1]#2{\@bsphack
  \@for\@citeb:=#2\do{%
    \edef\@citeb{\expandafter\@firstofone\@citeb}%
    \if@filesw\immediate\write\csname #1\endcsname{\string\citation{\@citeb}}\fi}%
  \@esphack}
\makeatother

\usepackage{enumitem}
\setlist{nosep, leftmargin=14pt}

\title{Flow Matching for Fluctuation-Based Fluorescence Super-Resolution}
\name{Alberto Cozzani$^{1,3}$ \qquad Giacomo Meanti$^{2}$ \qquad Laure Blanc-Féraud$^{3}$ \qquad Luca Calatroni$^{2,4}$}
\address{%
$^{1}$ Department of Mathematics, University of Bologna, Bologna, Italy \\
$^{2}$ MaLGa Center, DIBRIS, University of Genoa, Italy \\
$^{3}$ Université Côte d’Azur, CNRS, Sophia Antipolis, France\\
$^{4}$ MMS, Italian Institute of Technology (IIT), Italy
}
\begin{document}
\bstctlcite{IEEEexample:BSTcontrol}

\maketitle

\begin{abstract}
Fluctuation-based super-resolution techniques are an appealing alternative to single-molecule and structured-illumination microscopy, as they adapt to standard acquisition settings and exploit the independent stochastic fluctuations of fluorophores across a sequence of frames. However, most existing approaches lack an explicit model of the acquisition process and provide only relative intensity estimates, limiting quantitative analyses. In this work, we combine model-based and generative approaches to exploit the full stochastic information in the sequence, proposing a reconstruction framework built on flow matching and a double-Poisson forward model accounting for blinking, blur, under-sampling and noise.
The method alternates two steps: a flow matching model first learns the distribution of the acquired data; then, the sample and the background are jointly estimated by minimising a Kullback--Leibler divergence between observed and simulated distributions via regularised proximal gradient steps. On simulated and real Argolight data, the method resolves lines separated by 120~nm, outperforming SACD and eSRRF.
\end{abstract}

\section{Introduction}\label{sec:intro}

The spatial resolution of light microscopy is fundamentally limited by the diffraction limit, stemming directly from the physics of the optical system used for the acquisitions.
Single-molecule localisation microscopy (SMLM)~\cite{betzig2006imaging,rust2006sub} overcomes this limit by isolating and localising individual emitters, but requires photo-switchable fluorophores, high illumination intensities and thousands of sparse frames, resulting in long acquisitions, costly hardware and potential phototoxicity  which limits its use in live-cell imaging.
Fluctuation-based methods offer a cheaper alternative: they exploit the independent stochastic intensity fluctuations of fluorophores that naturally occur in standard widefield acquisitions, requiring minimal changes to the optical settings, being compatible with many fluorescent labels and relying on short captures.
Starting from SOFI~\cite{dertinger2009fast}, which reduces the effective PSF size through temporal correlations of pixel intensities, several methods have been proposed: SRRF~\cite{gustafsson2016fast} and its improved version eSRRF~\cite{laine2023high} apply correlation analysis to radiality maps, SACD~\cite{zhao2023enhanced} adds deconvolution steps to reduce out-of-focus contributions, while MUSICAL~\cite{agarwal2016multiple} and MSSR~\cite{torres2022extending} follow alternative statistical strategies.
A complementary line of work relies on sparse optimisation relying either on $\ell_1$ or $\ell_0$ relaxations, see SPARCOM~\cite{solomon2018sparsity} and COL0RME~\cite{stergiopoulou2022col0rme}, respectively.
Nonetheless, most of these methods exploit only selected statistics of the fluctuations (typically second-order) and provide relative intensities only, which limits quantitative downstream analyses.

In this work, we propose a physics-aware learning approach for image reconstruction which incorporates an explicit double-Poisson forward model of the fluorophore blinking, blur, under-sampling and noise.
Rather than relying on selected statistics, in analogy with FluoGAN ~\cite{cachia2023fluorescence}, the probability distribution of the observation sequence is here learned by means of a generative model, and then the forward parameters (i.e., high-resolution signal and out-of-focus background) are estimated with a tailored optimisation scheme so as to match the distribution of the observations.
For this, we leverage recent advances in flow matching (FM)~\cite{lipman2023flowmatchinggenerativemodeling} and diffusion models~\cite{ho2020denoising} to learn the data distribution and adapt to our task the losses described in~\cite{luo2023diff,meanti25}.
The result is a general framework for model-based inverse problems from i.i.d.\ noisy observations, which we apply to fluctuation microscopy with promising results on simulated and real data.

\section{Physics-informed distribution matching}\label{sec:method}

We assume that the (vectorised) high resolution ground-truth signal $\xtrue\in \Rn_{\geq 0}$ is observed through a sequence of undersampled, blurred and noisy observations $\bar{y}_t\in \Rm$, with $m<n$ and $t=1,\ldots,T$ corresponding to the images acquired by a microscope and containing information on the sample fluorophore intensity fluctuations across $T>1$ time-frames.
We model optical blur by means of a matrix $H \in \Rnn $ and include also a mixed Poisson-Gaussian noise modelling to describe both signal-dependent and readout noise.
The complete observation model reads
\begin{equation}\label{eq:obs}
    \bar{y}_t = \frac{1}{\alpha}\pois\left(\alpha\left(UH\frac{1}{\beta}\pois\left(\beta \xtrue \right)+\btrue \right)\right)+\eta_t
\end{equation}
where $\btrue\in \Rm_{\geq 0}$ is the background signal modelling out-of-focus signal, while
the inner stochastic Poisson process models the variability in time of the fluorophores, with the parameter $\beta>0$ depending on the underlying sample and photophysical properties \cite{omezzine2026mufasa}; $U \in \Rmn$ is used to describe the undersampling process of acquiring the signal on a coarser grid such that $n = r^2 m$ with $r>1$ being the super-resolution factor.
Finally, $\alpha>0$ and $\eta_t\sim\mathcal{N}(0, \sigma^2 I)$, for $\sigma>0$, describe the signal-dependent and Gaussian noise components.
Throughout the paper we assume that the acquisitions lie in a high photon count regime, where both $\beta \xtrue \gg 1$ and $\alpha (UH \beta^{-1}\pois(\beta \xtrue) + \btrue) \gg 1$ so that $\mathcal{P}(\lambda)\approx \mathcal{N}(\lambda,\lambda)$ is a valid approximation of the stochastic Poisson processes.

Given a sequence of frames $\ytrue = \{\bar{y}_t\}_{t=1}^T$, and having fixed/estimated both the optical and stochastic parameters $H,\alpha,\beta, \sigma$, we consider the problem of reconstructing the underlying high-resolution $\xtrue\in\Rn_{\geq 0}$ and background signal $\btrue\in\Rm_{\geq 0}$ from the observations.
Each observation in $\ytrue$ is assumed to be independently drawn from the same distribution $\ptrue$, which is realistic for independent acquisitions.
We solve this problem by defining a strategy for computing estimates $\xsim\approx \xtrue$ and $\bsim\approx \btrue$ so that having defined $\psim$ as
\begin{equation}
    y_t = \frac{1}{\alpha}\pois\left(\alpha\left(UH\frac{1}{\beta}\pois\left(\beta \xsim \right)+\bsim \right)\right)+\eta_t, ~~ y_t\sim\psim
    \label{eq:model_appr}
\end{equation}
we enforce $\psim$ to (approximately) match the unknown distribution $\ptrue$ \eqref{eq:obs}.
Following~\cite{luo2023diff,meanti25}, this requires a two-step approach:
\begin{enumerate}
    \item \emph{Learning the data distribution.} We approximate $\ptrue$ by means of a flow matching (FM) model computed using the independent samples in $\ytrue$.
    \item \emph{Distribution matching.} We seek $\xsim, \bsim$ such that an appropriate distance between $\ptrue$ (learned as in step 1.) and $\psim$ is minimised.
\end{enumerate}

\medskip

\noindent
\textbf{Learning the data distribution.}~~
The first step uses the observed data $\ytrue$ to train a FM model $v_{\theta}:\Rm\times[0,1]\to\Rm$ that maps samples distributed as Gaussian noise at time $\texttt{t} = 0$,
i.e.~$z^{(0)} \sim \mathcal{N}(0, I)$, to samples drawn from $\ptrue$ at time $\texttt{t} = 1$, i.e., $z^{(1)} \sim \ptrue$ through a window of time $\texttt{t}\in[0,1]$.
To do so, we consider the following FM loss~\cite{lipman2023flowmatchinggenerativemodeling}:
\begin{equation}\label{eq:fm-loss}
\mathcal{L}_{FM}(\theta) = \mathbb{E}_{\texttt{t},z^{(0)} ,z^{(1)}} \big\|v_{\theta} (z^{(\texttt{t})} , \texttt{t}) - (z^{(1)} - z^{(0)} )\big\|^2
\end{equation}
where $\texttt{t} \sim \mathcal{U}[0,1]$, $z^{(0)}$ and $z^{(1)}$ are independent, and $z^{(\texttt{t})} = (1-\texttt{t})\,z^{(0)} + \texttt{t}\,z^{(1)}$ is the linear interpolation between them.

\medskip
 
\noindent
\textbf{Estimating $x$ and $b$.}~~
We set $\xsim\in\Rn_{\geq 0}, \bsim\in\Rm_{\geq 0}$ as the optimisation variables, initialised at $\xsim_0$ and $\bsim_0$, respectively, and define $\ysim$ as a set of samples drawn from the forward model~\eqref{eq:model_appr} under the distribution $\psim= \psim(\xsim,\bsim)$.
We propose to compute an approximation of $(\xtrue,\btrue)\in\Rn_{\geq 0}\times \Rm_{\geq 0}$ by minimising the following integral Kullback-Leibler (KL) divergence between $\psim$ and $\ptrue$, measuring discrepancies between the two distributions through the FM process:
\begin{equation}
\mathrm{IKL}(\psim(\xsim, \bsim) \ \|  \ \ptrue)= \int_{\texttt{t}=0}^1 \mathrm{KL}( \psim^{(\texttt{t})}(\xsim,\bsim)\ \| \ \ptrue^{(\texttt{t})})d\texttt{t}
\label{eq:IKL}
\end{equation}
where $\ptrue^{(\texttt{t})}$ is the law of $\bar{y}^{(\texttt{t})}=(1-\texttt{t})\bar{y}^{(0)}+ \texttt{t} \bar{y}^{(1)}$, with $\bar{y}^{(0)}\sim\mathcal N(0,I)$, $\bar{y}^{(1)}\sim \ptrue$ and $\psim^{(\texttt{t})}$ is the law of $\ysim^{(\texttt{t})}$.
Given the joint dependence on $\xsim$ and $\bsim$ of $\psim$, we propose to alternate between (i) training an auxiliary FM model $v_\phi:\Rm\times[0,1] \to \Rm$ on $\ysim(\xsim, \bsim)$ to learn $\psim^{(\texttt{t})}$ by minimising~\eqref{eq:fm-loss}; (ii) update $\xsim$ and $\bsim$ by minimising the IKL divergence~\eqref{eq:IKL} using a proximal gradient algorithm. 
For this, we follow the derivation proposed by Diff-Instruct~\cite{luo2023diff}: the gradient of the IKL loss can be computed using the scores $s_\theta$ and $s_\phi$ of $\ptrue^{(\texttt{t})}$ and $\psim^{(\texttt{t})}(\xsim,\bsim)$ which can be derived from $v_\theta$ and $v_\phi$ respectively via $s(y,\texttt{t}) = (\texttt{t}v(y,\texttt{t}) - y)/(1-\texttt{t})$ so that:
{\small%
\begin{equation*}
    \nabla_{\xsim,\bsim} \mathrm{IKL}(\psim \ \|  \ \ptrue) = \int_{0}^1 \mathbb{E}[s_\theta(y^{(\texttt{t})}, \texttt{t})-s_\phi (y^{(\texttt{t})},\texttt{t})]^\top \cdot \nabla _{\xsim,\bsim} y ^{(\texttt{t})}d\texttt{t}
\end{equation*}%
}%
where the Gaussian approximation of the Poisson process can be used.
As $\ptrue$ depends on the unknown solution pair $(\xtrue,\btrue)$ only through the corruption process \eqref{eq:obs}, minimising the IKL alone may suffer from overfitting and not be able to identify $\xtrue$ and $\btrue$. Therefore, similarly to \cite{cachia2023fluorescence}, we incorporate model-based regularisation terms in the loss to promote sparsity on $\xsim$ and smoothness on $\bsim$, as well as a non-negativity constraint. For $\lambda_1,\lambda_2>0$, the final loss considered is
\begin{equation}
\min_{\xsim\ge0,~\bsim\geq 0}\ \mathrm{IKL}(\xsim, \bsim)+\lambda_1\|\xsim\|_1+\frac{\lambda_2}{2}\|\nabla\bsim\|^2,
\label{eq:loss}
\end{equation}
where $\text{IKL}(\cdot,\cdot)$ denotes the functional in \eqref{eq:IKL}. This problem is solved by iterating the following proximal gradient scheme:
\begin{mdframed}[linecolor=gray!60, linewidth=0.8pt]
\textbf{Proximal gradient algorithm \cite{attouch2013convergence}}
\scriptsize
\setlength{\abovedisplayskip}{8pt}
\begin{align*}
\xsim^{(k)} &= \max\!\big(0,\; \xsim^{(k-1)} - \tau\nabla_{\xsim}\mathrm{IKL}(\xsim^{(k-1)}, \bsim^{(k-1)}) - \tau\lambda_1\big), \\
\bsim^{(k)} &= \max\!\big(0,\; \bsim^{(k-1)} - \tau\nabla_{\bsim}\mathrm{IKL}(\xsim^{(k-1)}, \bsim^{(k-1)}) + \tau\lambda_2 \Delta \bsim^{(k-1)}\big),
\end{align*}
\end{mdframed}
\normalsize
Choosing $\tau<1/L$, with $L$ the Lipschitz constant of the gradient of the smooth part, guarantees monotone decrease of the objective and, under the Kurdyka--{\L}ojasiewicz property, convergence to a critical point~\cite{attouch2013convergence}. However, as $L$ is not available and $\nabla\mathrm{IKL}$ is estimated stochastically via $v_\theta$ and $v_\phi$, step sizes are tuned empirically.
 Note that the $\xsim$-update corresponds to the computation of $\operatorname{prox}_{\tau(\lambda_1\|\cdot\|_1 + \iota_{\ge 0})}$ which is the composition of a soft-thresholding and a standard projection due to both the structure of the (box) constraint and the separability of the $\ell_1$ norm.

\section{Numerical Results}\label{sec:results}
We evaluate the capability of the proposed approach to super-resolve widefield fluorescence microscopy images, achieving resolutions smaller than the diffraction limit evaluated thanks to Argolight SIM slide. Our first results are on data simulated by means of the double-Poisson forward model~\eqref{eq:obs}. Next, we show the robustness of the proposed approach on real images of an ARGO-SIM pattern~\cite{laine2023high}.

\medskip

\noindent
\textbf{Simulated Argolight data.}~~
We simulate $T=500$ frames of an ARGO-SIM calibration pattern consisting of 8 groups of 4 parallel lines. The centre-to-centre separation distance $d$ between the two inner lines increases progressively in steps of \SI{30}{\nano\meter} starting from \SI{120}{\nano\meter}.
Fluorophore emission along the lines is modelled pixel-wise using a Poisson distribution with $\beta=1$ and corrupted according to~\eqref{eq:obs} with an under sampling factor $r=6$, Gaussian blur with $\mathrm{FWHM}=\SI{311}{\nano\meter}$, and Poisson-Gaussian noise with $\alpha=12, \sigma=5$.
The corrupted frames have a final pixel size of \SI{100}{\nano\meter}.

\begin{figure}[htbp]
    \centering
    \def\figcolsep{1.25}
    \def\figzoomy{-3.1}
    \def\figplotsep{0.2cm} 
    \def\figploth{6.1cm}   

    \begin{tikzpicture}[
        spy using outlines={shape=rectangle, size=1.1cm, lens={scale=3.5}, thick, yellow}
    ]
        \node[inner sep=0pt] (imgCorr) at (0, 0) {
            \includegraphics[angle=90, width=1.2cm]{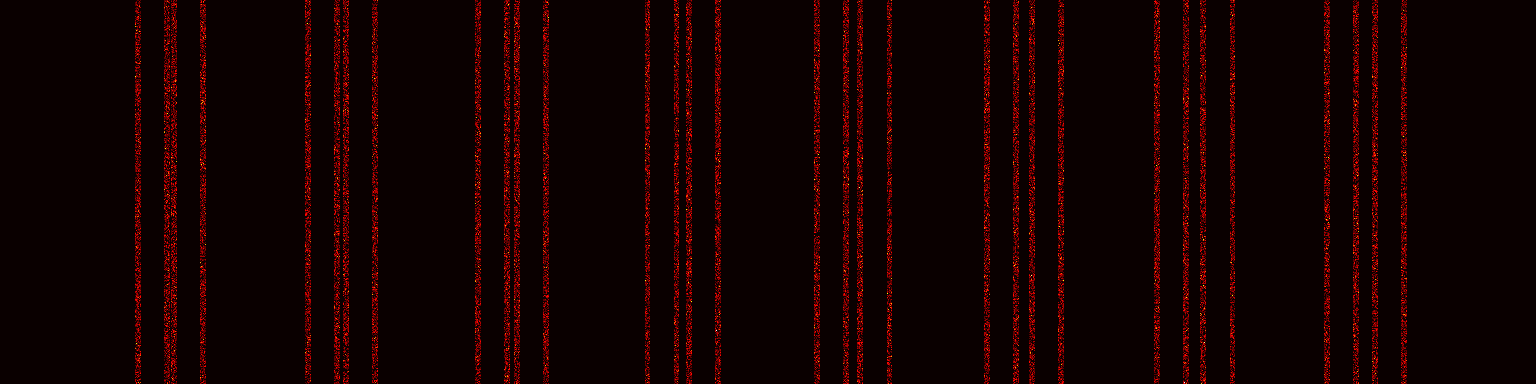}};
        \node[inner sep=0pt] (imgOurs) at (\figcolsep, 0) {
            \includegraphics[angle=90, width=1.2cm]{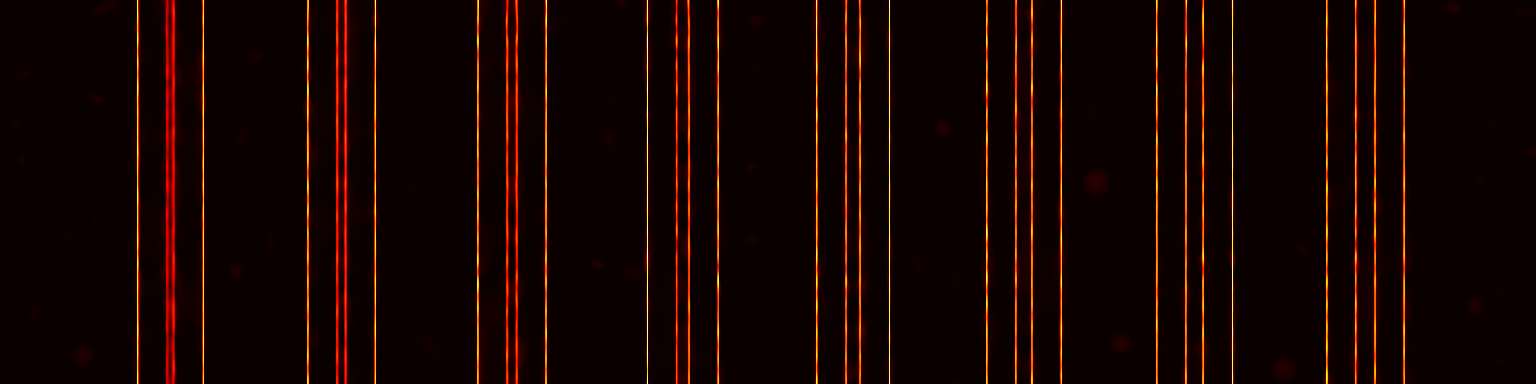}};
        \node[inner sep=0pt] (imgSRRF) at ({2*\figcolsep}, 0) {
            \includegraphics[angle=90, width=1.2cm]{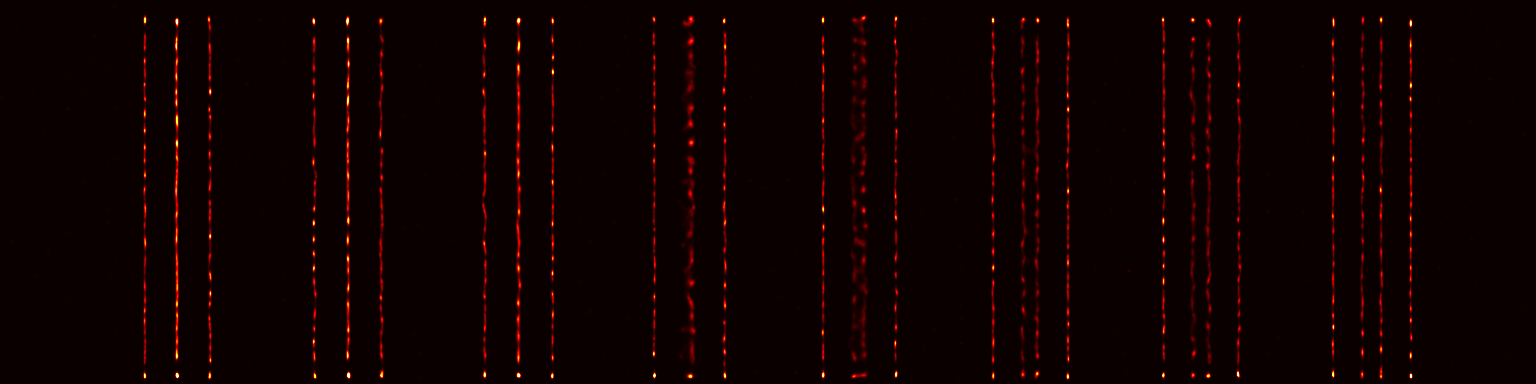}};
        \node[inner sep=0pt] (imgeSRRF) at ({3*\figcolsep}, 0) {
            \includegraphics[angle=90, width=1.2cm]{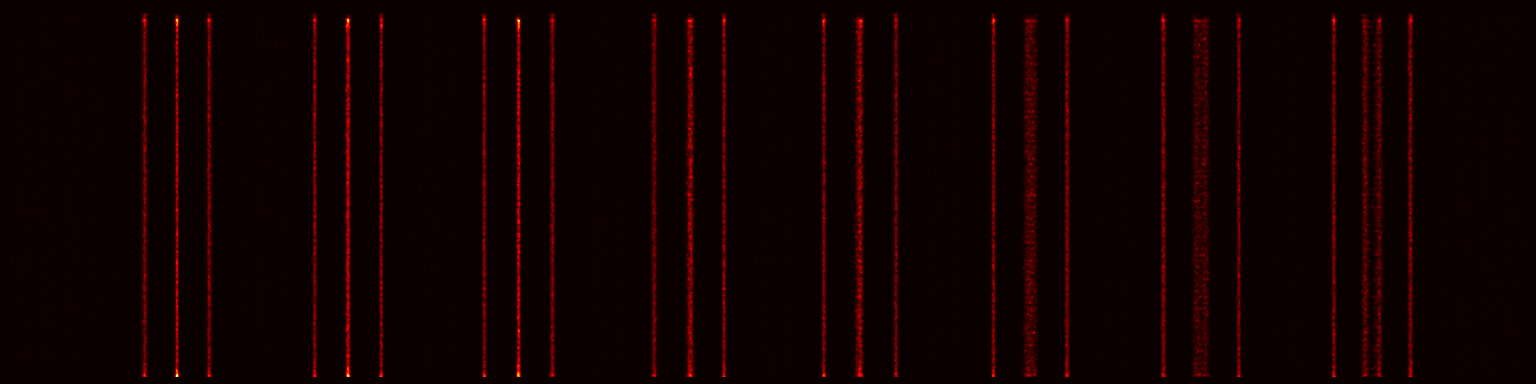}};

        \node[above=0.1cm, font=\fontsize{9}{1}\bfseries] at (imgCorr.north) {GT};
        \node[above=0.1cm, font=\fontsize{9}{1}\bfseries] at (imgOurs.north) {Ours};
        \node[above=0.1cm, font=\fontsize{9}{1}\bfseries] at (imgSRRF.north) {SACD};
        \node[above=0.1cm, font=\fontsize{9}{1}\bfseries] at (imgeSRRF.north) {eSRRF};

        \spy on ([yshift=-1.86cm]imgCorr.center) in node [anchor=center] at (0, \figzoomy);
        \spy on ([yshift=-1.86cm]imgOurs.center) in node [anchor=center] at (\figcolsep, \figzoomy);
        \spy on ([yshift=-1.86cm]imgSRRF.center) in node [anchor=center] at ({2*\figcolsep}, \figzoomy);
        \spy on ([yshift=-1.86cm]imgeSRRF.center) in node [anchor=center] at ({3*\figcolsep}, \figzoomy);

        \newcommand{\fsd}{9}
        \node[inner sep=0pt, anchor=north west] (plots)
            at ([xshift=\figplotsep]imgeSRRF.north east |- imgCorr.north)
            {\includegraphics[height=\figploth]{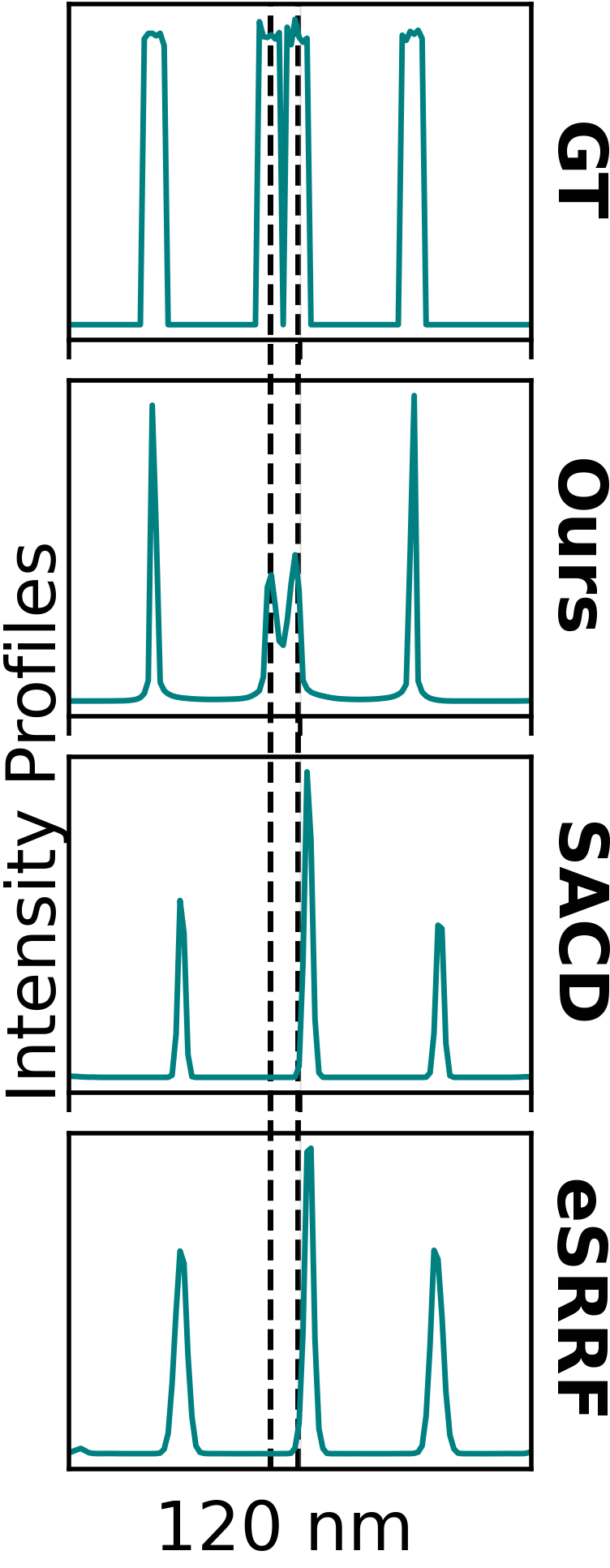}};
        \node[anchor=east, font=\fontsize{\fsd}{5}\selectfont] at ([xshift=-0.05cm, yshift=-1.86cm]imgCorr.west) {120};
        \node[anchor=east, font=\fontsize{\fsd}{5}\selectfont] at ([xshift=-0.05cm, yshift=-1.32cm]imgCorr.west) {150 };
        \node[anchor=east, font=\fontsize{\fsd}{5}\selectfont] at ([xshift=-0.05cm, yshift=-0.78cm]imgCorr.west) {180 };
        \node[anchor=east, font=\fontsize{\fsd}{5}\selectfont] at ([xshift=-0.05cm, yshift=-0.25cm]imgCorr.west) {210 };
        \node[anchor=east, font=\fontsize{\fsd}{5}\selectfont] at ([xshift=-0.05cm, yshift=0.27cm]imgCorr.west) {240 };
        \node[anchor=east, font=\fontsize{\fsd}{5}\selectfont] at ([xshift=-0.05cm, yshift=0.80cm]imgCorr.west) {270 };
        \node[anchor=east, font=\fontsize{\fsd}{5}\selectfont] at ([xshift=-0.05cm, yshift=1.35cm]imgCorr.west) {300 };
        \node[anchor=east, font=\fontsize{\fsd}{5}\selectfont] at ([xshift=-0.05cm, yshift=1.9cm]imgCorr.west) {330};
        \node[anchor=east, font=\fontsize{\fsd}{5}\selectfont] at ([xshift=-0.05cm, yshift=2.25 cm]imgCorr.west) {d (nm)};
    \end{tikzpicture}
    \caption{Simulated Argolight comparison. \textbf{Left:} Reconstructions (left to right: ground truth, ours, SACD and eSRRF). The yellow boxes highlight the magnified insets corresponding to the $120$ nm line separation. \textbf{Right:} Normalised intensity profiles for the $d=120$ nm line group.}
    \label{fig:argolight_comparison}
\end{figure}

\begin{figure}[t]
    \centering
    \includegraphics[width=0.99\linewidth]{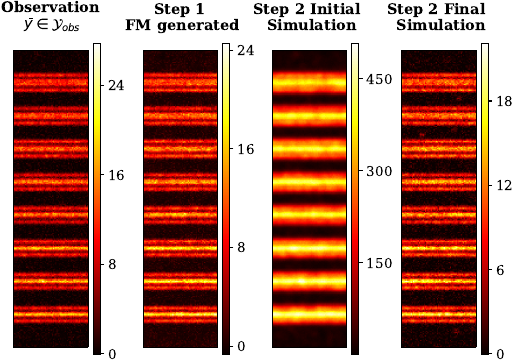}
    \caption{Evolution of the observations during the two-step reconstruction process. From left to right: an acquired frame $\bar{y}_t\in\ytrue$, a sample generated by the flow matching model after Step 1, and simulated observations at the initialisation and final stages of Step 2.}
    \label{fig:corruptions}
\end{figure}

\noindent To quantitatively assess line separation, we computed the 1D intensity profiles of the reconstructions by summing pixel intensities along the direction of the lines and normalising them in the range $[0,1]$. We evaluated our method on the simulated data and found optimal performance with regularisation weights $\lambda_1 = 0.05$ and $\lambda_2 = 40$ in terms of the peak-to-valley ratio of the profiles. We compare our results with those obtained with SACD~\cite{zhao2023enhanced} (using 20 initial and 100 final LR iterations and with the exact PSF) and eSRRF~\cite{laine2023high} (with radius $R=1$ and sensitivity $S=1$). 
As illustrated in Fig.~\ref{fig:argolight_comparison}, our method successfully resolves the two inner lines down to the finest tested separation of \SI{120}{\nano\meter}, as also confirmed by the intensity profiles plot.  In contrast, under the tested configurations, eSRRF and SACD are only capable of resolving line pairs separated by at least \SI{300}{\nano\meter} and \SI{240}{\nano\meter}, respectively. 
Fig.~\ref{fig:corruptions} shows the outputs at the different steps of our method, comparing observations from $\ytrue$ with those generated at Step~1, at the start of Step~2 and at convergence of Step~2. The strong visual alignment between $\ytrue$ and the samples drawn from the distribution $\ptrue$ learned at Step~1 confirms that FM accurately models the degradation process. Furthermore, the similarity between $\ytrue$ and the simulated degradation $\ysim$ at the end of Step~2 validates that the variational distance minimization yields physically consistent results.

\medskip

\noindent\textbf{Real ARGO-SIM.}~~ Raw images, kindly provided by the authors of \cite{laine2023high}, were acquired on a Carl Zeiss ELYRA PS.1 microscope equipped with an $\alpha$Plan-Apochromat 100×/1.46 oil-immersion objective and an Andor EMCCD 887 1K camera \cite{laine2023high}. A $128\times128$ pixel region of interest of the gradually spaced line pattern on the Argo-SIM calibration slide was recorded at 86 frames per second, yielding a sequence of $T=1000$ frames. The structure is identical to the one described above. The FWHM was estimated at 167 nm, while the noise parameters $\alpha$ and $\sigma$ were derived through mean-variance curve fitting.

\begin{figure}
    \centering
            \includegraphics[width=\linewidth]{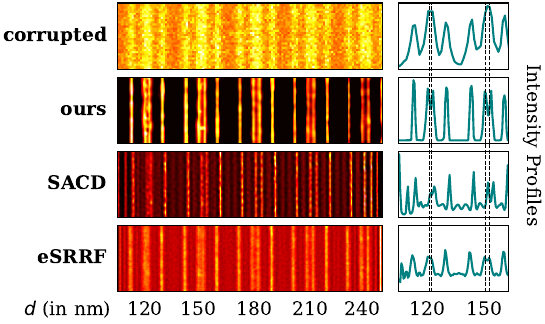}
    \caption{Real Argolight comparison. \textbf{Left:} Reconstructions (top to bottom: corrupted, ours, SACD, eSRRF). \textbf{Right:} Normalised intensity profiles for the $d=120$ nm and $d=150$ nm line groups.}
    \label{fig:argolight_czi_comparison}
\end{figure}

We evaluated our method on these corrupted images using an upsampling factor of $2$.  Setting the regularisation weights to $\lambda_1 = 0.80$ and $\lambda_2=60$ yielded optimal results in terms of the peak-to-valley ratio of the 1D intensity profiles, allowing us to resolve lines down to the minimum considered separation of $120$ nm. Fig.~\ref{fig:argolight_czi_comparison} benchmarks our reconstruction against SACD and eSRRF. We performed SACD with $20$ initial and $100$ final LR iterations, setting the microscope numerical aperture and emission wavelength taken from the acquisitions metadata. eSRRF was run with $R=1$ and $S=2$, following the recommendations of~\cite{laine2023high}. While eSRRF and SACD struggle to provide a clear and distinctly separated 1D intensity profile at $120$ nm, our method produces visually sharper, distinctly isolated peaks with higher contrast and a significantly suppressed background noise. Furthermore, although SACD results in sharper lines than eSRRF, it introduces edge artifacts, thus hallucinating spurious lines at the boundaries.

\section{Conclusions}\label{sec:conclusions}
In this paper, we presented a physics-informed generative approach for fluctuation-based fluorescence microscopy super-resolution, combining flow matching techniques  with model-based regularisation taking advantage of a double-Poisson problem describing the physical behaviour of the fluorophores and the whole image formation process. Experiments on both simulated and real Argolight data demonstrate promising resolving capabilities compared with the existing fluctuation-based approaches. Future work shall focus on the extension of the framework considered to accommodate non-static scenarios and photo-bleaching, alongside further validation on live-cell biological data. The code of the experiments reported in this work is available at: \url{https://github.com/cozzalberto/flow_fluo}

{\small \section{Acknowledgments}
The authors thank Audrey Salles from Institut Pasteur for acquiring and curating the ARGO-SIM dataset used in this study, and the authors of~\cite{laine2023high} for kindly sharing it. This work has been supported by the French government, through the 3IA Cote d’Azur Investments in the ANR project ANR-23-IACL-0001.
LC and GM are supported by the ERC starting grant MALIN (101117133). This work represents only the view of the authors.}
{\small \section{Compliance with Ethical Standards}
This study was performed using simulated data and physical calibration patterns. No human subjects or animal experiments were involved; therefore, ethical approval was not required.}
\ninept
\bibliographystyle{IEEEtran}
\bibliography{strings,refs}

@article{attouch2013convergence,
  author  = {Attouch, H{\'e}dy and others},
  title   = {Convergence of descent methods for semi-algebraic and tame problems: proximal algorithms, forward--backward splitting, and regularized {G}auss--{S}eidel methods},
  journal = {Math. Program.},
  year    = {2013}
}

@article{omezzine2026mufasa,
  author  = {Omezzine, Wessim and others},
  title   = {{MUFASA}: a continuous-time stochastic framework for realistic fluorescence microscopy simulation},
  journal = {Commun. Biol.},
  year    = {2026}
}

@article{betzig2006imaging,
  author  = {Betzig, Eric and others},
  title   = {Imaging intracellular fluorescent proteins at nanometer resolution},
  journal = {Science},
  year    = {2006}
}

@article{rust2006sub,
  author  = {Rust, Michael J. and others},
  title   = {Sub-diffraction-limit imaging by stochastic optical reconstruction microscopy ({STORM})},
  journal = {Nat. Methods},
  year    = {2006}
}

@inproceedings{meanti25,
  author    = {Meanti, Giacomo and others},
  title     = {Unsupervised imaging inverse problems with diffusion distribution matching},
  booktitle = {ICCV},
  year      = {2025}
}

@article{laine2023high,
  author  = {Laine, Romain F. and others},
  title   = {High-fidelity {3D} live-cell nanoscopy through data-driven enhanced super-resolution radial fluctuation},
  journal = {Nat. Methods},
  year    = {2023}
}

@article{gustafsson2016fast,
  author  = {Gustafsson, Nils and others},
  title   = {Fast live-cell conventional fluorophore nanoscopy with {ImageJ} through super-resolution radial fluctuations},
  journal = {Nat. Commun.},
  year    = {2016}
}

@inproceedings{luo2023diff,
  author    = {Luo, Weijian and others},
  title     = {Diff-{I}nstruct: A universal approach for transferring knowledge from pre-trained diffusion models},
  booktitle = {NeurIPS},
  year      = {2023}
}

@inproceedings{lipman2023flowmatchinggenerativemodeling,
  author    = {Lipman, Yaron and others},
  title     = {Flow matching for generative modeling},
  booktitle = {ICLR},
  year      = {2023}
}

@article{dertinger2009fast,
  author  = {Dertinger, Thomas and others},
  title   = {Fast, background-free, {3D} super-resolution optical fluctuation imaging ({SOFI})},
  journal = {PNAS},
  year    = {2009}
}

@article{torres2022extending,
  author  = {Torres-Garc{\'\i}a, Esley and others},
  title   = {Extending resolution within a single imaging frame},
  journal = {Nat. Commun.},
  year    = {2022}
}

@article{zhao2023enhanced,
  author  = {Zhao, Weisong and others},
  title   = {Enhanced detection of fluorescence fluctuations for high-throughput super-resolution imaging},
  journal = {Nat. Photon.},
  year    = {2023}
}

@article{agarwal2016multiple,
  author  = {Agarwal, Krishna and Mach{\'a}{\v{n}}, Radek},
  title   = {Multiple signal classification algorithm for super-resolution fluorescence microscopy},
  journal = {Nat. Commun.},
  year    = {2016}
}

@article{solomon2018sparsity,
  author  = {Solomon, Oren and others},
  title   = {Sparsity-based super-resolution microscopy from correlation information},
  journal = {Opt. Express},
  year    = {2018}
}

@inproceedings{ho2020denoising,
  author    = {Ho, Jonathan and others},
  title     = {Denoising diffusion probabilistic models},
  booktitle = {NeurIPS},
  year      = {2020}
}

@article{cachia2023fluorescence,
  author  = {Cachia, Mayeul and others},
  title   = {Fluorescence image deconvolution microscopy via generative adversarial learning ({FluoGAN})},
  journal = {Inverse Probl.},
  year    = {2023}
}

@article{stergiopoulou2022col0rme,
  author  = {Stergiopoulou, Vasiliki and others},
  title   = {{COL0RME}: Super-resolution microscopy based on sparse blinking/fluctuating fluorophore localization and intensity estimation},
  journal = {Biol. Imaging},
  year    = {2022}
}

\end{document}